\documentclass[11pt,twocolumn,tighten]{aastex631}
\usepackage{amsmath}
\usepackage{url}

\newcommand{\p}{\partial}
\newcommand{\f}{\frac}

\begin{document}
\title{A Deep Search for a Strong Diffuse Interstellar Band in the Circumgalactic Medium}
\correspondingauthor{Chih-Yuan Chang, Ting-Wen Lan}
\email{ccychihyuan@gmail.com, twlan@ntu.edu.tw}
\author{Chih-Yuan Chang}
\affiliation{Department of Physics, National Taiwan University, No. 1, Sec. 4, Roosevelt Rd., Taipei 10617, Taiwan}
\affiliation{Institute of Astronomy and Astrophysics, Academia Sinica, No. 1, Sec. 4, Roosevelt Rd., Taipei 10617, Taiwan}

\author[0000-0001-8857-7020]{Ting-Wen Lan}
\affiliation{Graduate Institute of Astrophysics, National Taiwan University, No. 1, Sec. 4, Roosevelt Rd., Taipei 10617, Taiwan}
\affiliation{Department of Physics, National Taiwan University, No. 1, Sec. 4, Roosevelt Rd., Taipei 10617, Taiwan}
\affiliation{Institute of Astronomy and Astrophysics, Academia Sinica, No. 1, Sec. 4, Roosevelt Rd., Taipei 10617, Taiwan}

\begin{abstract}
We investigate the absorption signals of a strong diffuse interstellar band, DIB$\lambda4430$, in the circumgalactic medium (CGM) traced by MgII absorption lines. To this end, we make use of approximately 60,000 MgII absorption line spectra within $0.4<z<1.0$ compiled from the Sloan Digital Sky Surveys and obtain composite spectra with uncertainties for absorption line measurements being a few $\rm m\AA$. By using MgII absorption strength and dust reddening relation from the literature, we measure the DIB$\lambda4430$ absorption strength as a function of $\rm E(B-V)$ in the CGM, and compare the Milky Way DIB$\lambda4430$ - $\rm E(B-V)$ relation extrapolated down to the CGM E(B-V) region. Our results show no detectable signals of DIB$\lambda4430$ across the entire $\rm E(B-V)$ range in the CGM traced by MgII absorption lines. This lack of detection of DIB$\lambda4430$ in the CGM is inconsistent with the Milky Way signals by $\sim 5 \, \sigma$, indicating that the factors associated with different environments affect the abundance of the DIB$\lambda4430$ carrier. 
\end{abstract}

\section{Introduction}

Diffuse interstellar bands (DIBs) are weak absorption features that have been detected ubiquitously in the Milky Way and other galaxies \citep[e.g.,][]{Jenniskens1994,Hobbs2008,Hobbs2009,Cordiner2011,Baron2015,Lan2015}. The first discovery of DIBs dates back to 1919 \citep{Heger1922} with the detection of two DIBs ($\lambda {5780}\ \text{and}\ \lambda {5797}$) in the stellar spectra within our Milky Way, which were later found to be interstellar in origin \citep{Merrill1934,Merrill1938}.
In total, over 500 DIBs have been found \citep[e.g.,][]{Hobbs2008,Hobbs2009}. It is speculated that the sources of these absorption features are large organic molecules such as polycyclic aromatic hydrocarbons (PAHs) \citep[e.g.,][]{Crawford1985,Leger1985}, fullerenes \citep[e.g.,][]{Leger1988}, or fulleranes \citep[e.g.,][]{Webster1993}. In 2015, laboratory studies confirmed $C_{60}^+$ to be the carrier of two DIBs, $\lambda9632$ and $\lambda9577$ \citep{Campbell2015}, and further studies \citep[e.g.][]{Campbell2016,Cordiner2017,Campbell2018,Cordiner2019} have  identified three more DIBs as arising from $C^+_{60}$ transitions within the interstellar medium \citep{Linnartz2020}. These results demonstrate that large complex molecules similar to $C_{60}$ are possible carriers for other DIBs. 

To assist the identification of DIB carriers, one can  detect and characterize the properties of DIBs in different environments \citep[e.g.,][]{Jenniskens1994,Galazutdinov2000,vanLoon2013,Welty2014,Baron2015,Lan2015,Elyajouri2017,Schultheis2023}. 
Previous studies have mostly focused on probing DIBs in the relatively gas-rich and dusty environments in our Milky Way, where DIB absorptions have been found to be stronger \citep[e.g.][]{Baron2015,Lan2015}. There are also explorations of DIBs in the interstellar medium (ISM) of other galaxies \citep[e.g.,][]{Cordiner2014}. For example, DIBs have been detected in Large and Small Magellanic clouds \citep[e.g.,][]{Welty2006, Bailey2015} and local starburst galaxies \citep[e.g.,][]{Heckman2000,Welty2014b}. In addition, the abundances of DIBs have been investigated in quasar absorption line systems with high neutral hydrogen column densities \citep[DLAs, damped Ly$\alpha$ system,][]{York2006,Lawton2006} and with high metal column densities (CaII) \citep{Ellison2008}. However, due to their weak absorption strengths, the content of DIBs in low-density environments, such as in the circumgalactic medium \citep[CGM,][]{Tumlinson2017}, has not been explored. 

In this work, we make use of the large spectroscopic dataset provided by the Sloan Digital Sky Surveys (SDSS) \citep{York2000} and perform a deep search for one of the strongest DIBs, DIB$\lambda4430$ \citep[e.g.,][]{Walker1963, Snow2002a, Snow2002}, in the CGM. We use MgII absorption lines detected in quasar spectra as a tracer of the cool CGM ($T\sim10^{4}$ K) \citep[e.g.,][]{Lan2018, Lan2020, Anand2021} and obtain high signal-to-noise (S/N) composite spectra by combining tens of thousands of individual absorption line spectra. This allows us to probe weak DIBs absorbing $\sim0.1\%$ of the continuum.

The structure of the paper is as follows. In Section~\ref{section_2}, we  detail the data analysis process, including the data and the methodology of obtaining the composite spectra. We summarize the results and discuss the implications of our findings in Section~\ref{section_3}. We conclude this work in Section \ref{section_4}. Throughout the paper, we use vacuum wavelength.

\section{Data Analysis}
\label{section_2}

\subsection{Composite spectra for DIB$\lambda 4430$}
To probe DIB$\lambda4430$ in the CGM traced by MgII absorption lines, we use a large MgII absorption catalog\footnote{\url{https://wwwmpa.mpa-garching.mpg.de/SDSS/MgII/}} compiled by \cite{Anand2021} from the Sloan Digital Sky Survey DR16 dataset \citep{Lyke2020}. In total, the catalog contains $\sim160,000$ MgII absorbers with redshifts ranging from 0.36 to 2.28 \citep{Anand2021}. In order to detect DIB$\lambda4430$ in the SDSS optical spectra, we select MgII absorbers with redshifts between $\rm 0.4<z<1.0$. This yields 60,650 MgII absorbers with a median redshift of $\sim0.7$.

To further combine individual absorption spectra, we follow a similar procedure as \citet{Zhu2014}.  We first estimate the intrinsic spectral energy distributions (SEDs) of the quasars by using a set of quasar eigenspectra obtained by applying non-negative matrix factorization technique (implemented by \citet{Zhu2016}) to the SDSS quasar spectra from \citet{Zhu2013}. 
The original spectra are then normalized by the NMF-reconstructed quasar SEDs. We further apply a median filter with 71 pixels to 
the normalized spectrum to obtain a median continuum for capturing small-scale fluctuation with the pixels between 4420 and 4440 $\rm \AA$ being masked. 
The final normalized spectrum is the ratio between the normalized spectrum and the median continuum. All the normalized spectra are shifted into the rest-frame of the MgII absorbers and projected onto a common wavelength grid with the SDSS spectral resolution. 
We combine individual spectra with a median estimator to avoid effects of outliers in the spectra. We note that this method has been used to detect weak spectral features in previous works \citep[e.g.,][]{Zhu2013b, Zhu2014, Lan2017, Lan2018, Ng2024}. 
Finally, we obtain high signal-to-noise composite spectra as a function of the rest equivalent widths of MgII absorption lines ($W_0^{\lambda2796}$). Table~\ref{table_1} summarizes the number of individual spectra used in the composite spectra. 

\begin{table}[h]
\centering
\caption{Number of spectra used in each composite spectrum as a function of $W_0^{\lambda2796}$ 
}
\begin{tabular}{c | c } 
 \hline
  $W_0^{\lambda2796}$ [\AA] & Number of Spectra\\ [0.5ex] 
 \hline
 
$0.40<W_0^{\lambda2796}<0.66$ & 7641 \\ 
$0.66<W_0^{\lambda2796}<1.09$ & 18453 \\
$1.09<W_0^{\lambda2796}<1.82$ & 22640 \\
$1.82<W_0^{\lambda2796}<3.07$ & 10959 \\ 
$3.07<W_0^{\lambda2796}$ & 957 \\[1ex] 
 \hline
\end{tabular}

\label{table_1}
\end{table}

To measure possible weak $\rm DIB\lambda4430$ signals from the composite spectra, we fit a Gaussian profile with a fixed line width $\sigma=6.9$ \AA{} and a fixed central wavelength at 4429.9 \AA, the best-fit parameters of Milky Way sightlines from \citet{Lan2015}. The uncertainty of absorption strength is estimated via bootstrapping the sample 500 times.

\begin{figure}[ht]
    \centering
    \includegraphics[width=0.48\textwidth]{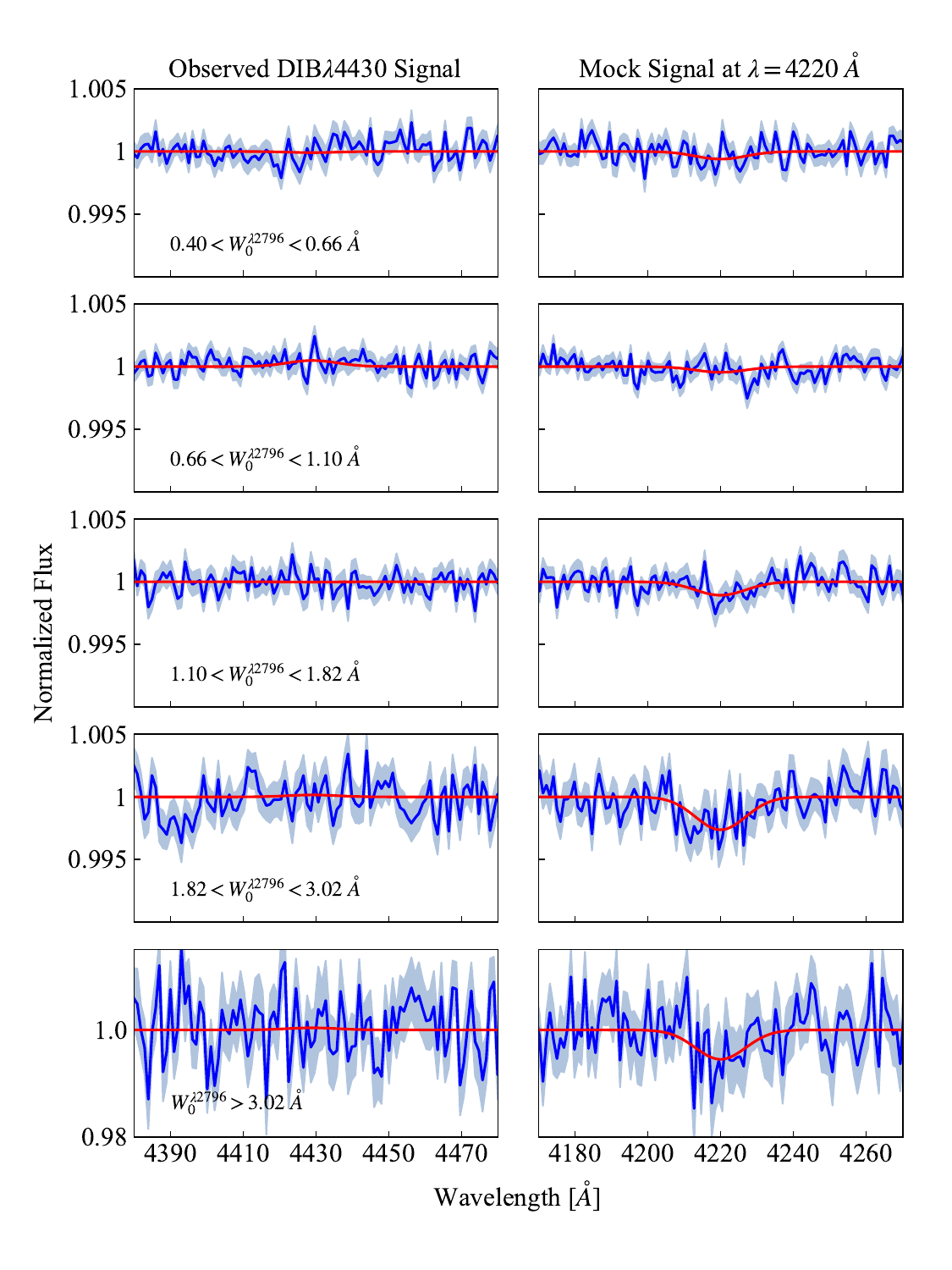}
    \caption{Composite spectra as a function of $W_0^{\lambda2796}$. The blue solid lines show the composite spectra with bootstrap uncertainties indicated by the shaded bands. The red solid lines are the best-fit absorption profiles.
    \emph{Left:} observed spectral regions around DIB$4430$. \emph{Right:} spectral regions around 4220 $\rm \AA$ with synthetic absorption signals recovered.} 
    \label{MgBins}
\end{figure}

\subsection{Dust reddening of MgII absorbers}
In order to compare the $\rm DIB\lambda4430$ signals in the CGM traced by MgII absorbers and the $\rm DIB\lambda4430$ signals observed in the Milky Way with a given dust content along the lines of sight, we use the $W_0^{\lambda2796}$ and dust reddening E(B-V) based on \citet{Menard2012} and \citet{Menard2008} with
\begin{equation}
    E(B-V)_{obs}\simeq \frac{E(g-i)_{obs}}{1.55}\simeq\frac{0.017}{1.55}\times\bigg(\frac{W_0^{\lambda2796}}{1 \rm \AA}\bigg)^{1.6},
    \label{dust extinction}
\end{equation}
\begin{equation}
   E(B-V)_{rest} = E(B-V)_{obs}\times(1+z)^{-1.2}
\end{equation}
where $E(g-i)_{obs}$ is the dust reddening with SDSS g and i bands in observed frame, $E(B-V)_{obs}$ is the estimated reddening with SMC like extinction curve \citep{York2006,Menard2012}, and $E(B-V)_{rest}$ is the reddening in the rest-frame. The above parameter values are adopted from \citet{Menard2012} and \citet{Menard2008} based on the color excess signals induced by MgII absorbers. 

For the DIB$\lambda4430$ signals in the Milky Way, we use the DIB$\lambda4430$ absorption strength based on the relation observed in the Milky Way from \citet{Lan2015}:
\begin{equation}
    W_{0}^{\lambda4430}(E(B-V))=1.22 (E(B-V)_{rest})^{0.89}  \, \rm \AA.
    \label{DIB_dust}
\end{equation}
We note that the average E(B-V) value for MgII absorbers with $\rm W_0^{\lambda2796}\sim1 \, \AA$ is $\sim0.005$ mag, which is lower than a typical Milky Way sightline. Therefore, the expected MW DIB signals are extrapolated values based on the best-fit relation from $E(B-V)>0.04 \, \rm mag$ sightlines \citep{Lan2015}.

Combining E(B-V) in MgII absorbers and the correlation between DIB$\lambda4430$ and E(B-V) in our Milky Way, we compare DIB signals in the CGM and the expected signals in the Milky Way with the same E(B-V) value. 

\section{Results and Discussion}
\label{section_3}

\begin{figure*}[ht]
    \centering
    \includegraphics[width=0.75\textwidth]{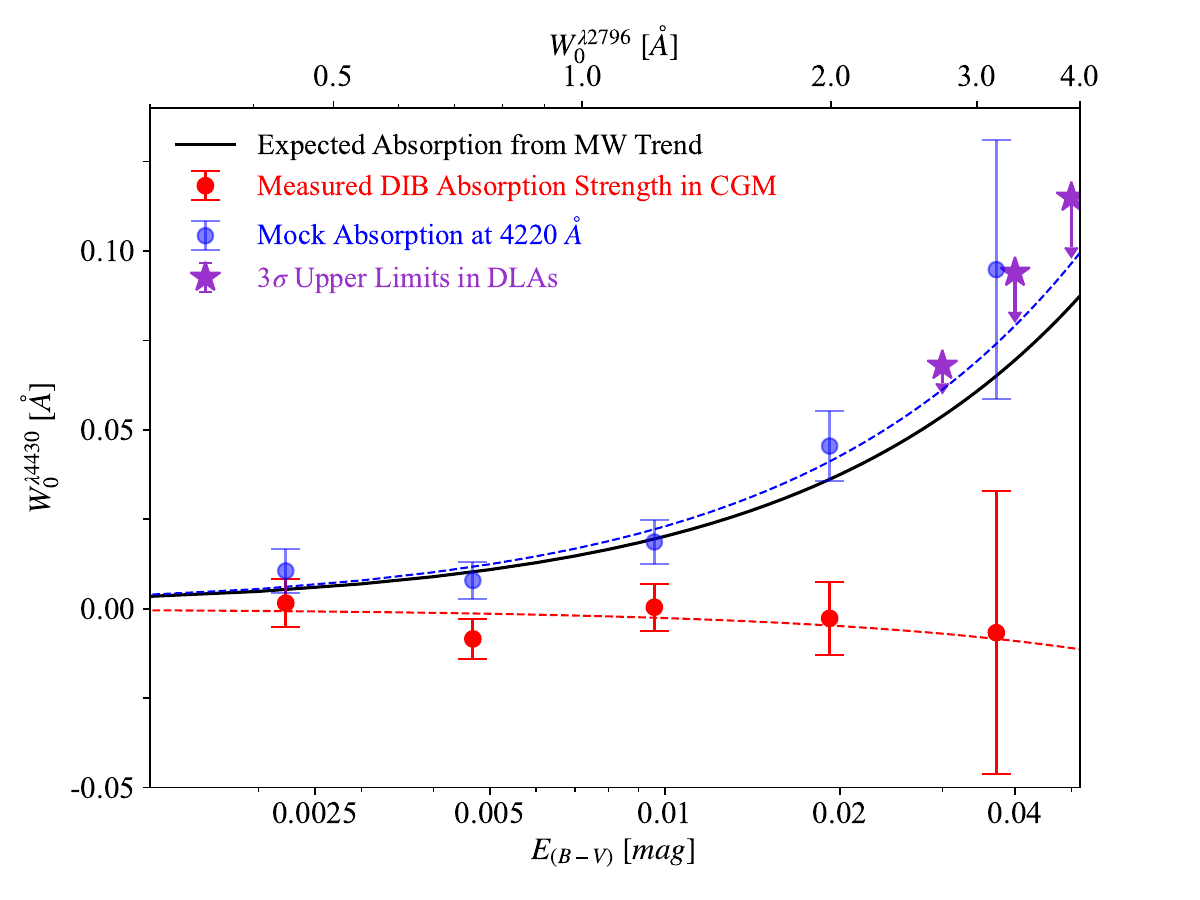}
    \caption{DIB$\lambda4430$ absorption strengths as a function of $W^{\lambda2796}_{0}$ and $E(B-V)$. The red data points show the measured absorption strengths at DIB4430 and the blue data points show the measured absorption strengths for the synthetic absorption profiles at 4220 \AA{}. The corresponding color dashed lines show the best-fit power laws. The black solid line shows the Milky Way DIB$\lambda4430$ - E(B-V) relation which is recovered from the synthetic signals (blue line). The purple stars show 3$\sigma$ upper limits derived from damped Lyman-alpha systems \citep{Lawton2006}.}
    \label{EvsO}
\end{figure*}

The composite spectra are shown in Figure~\ref{MgBins}. The left panels show the spectral regions around DIB$\lambda4430$. The solid lines are the composite spectra with bootstrapping uncertainties indicated by the shaded regions. The red solid lines show the best-fit profiles for measuring the possible absorption strengths of DIB$\lambda4430$. 
The rest equivalent width of DIB$\lambda4430$ as a function of MgII absorption and E(B-V) are shown in Figure~\ref{EvsO}. We do not detect any DIB$\lambda4430$ absorption throughout the range of E(B-V) traced by MgII absorbers. 
To quantify the overall trend, we fit our measured $W^{\lambda4430}_{0}$ as a function of E(B-V) with 
\begin{equation}
    W^{\lambda4430}_{0} = A\times E(B-V)^{0.89},
\end{equation}
where the power index 0.89 is the expected value based on the Milky Way relation (Eq. 3). 
The best-fit value of $A$ is $-0.16 \pm 0.23$ being consistent with no detection. 
This non-detection result deviates from the expected signals based on the $W^{\lambda4430}_{0}-E(B-V)$ relation (black line in Figure~\ref{EvsO}) from the Milky Way sightlines \citep{Lan2015}. 

To further test the results, we insert mock DIB absorption lines at $4220 \rm \, \AA$ in the rest frame of MgII absorbers into the original SDSS quasar spectra with the strengths following the $W^{\lambda4430}_{0}-E(B-V)$ relation in the Milky Way (Eq. 3). This spectral region is chosen for its lack of absorption features. We run the same pipeline, which includes NMF normalization and median filter, to obtain the median composite spectra. The right panels of Figure~\ref{MgBins} show the composite spectra around $4220 \rm \, \AA$. The red solid lines show the best-fit absorption profiles, recovering the input signals. The blue data points in Figure~\ref{EvsO} are the recovered signals, which are consistent with input signals (black solid line). 
We also fit the mock signals with Eq. 4 and obtain the best-fit value of $A_{mock}$ being $1.39 \pm 0.22$, which is consistent with the input relationship (Eq. 3). 
Based on the non-detection of DIB$\lambda4430$ and the corresponding uncertainty, our result shows that the $W^{\lambda4430}_{0}$-E(B-V) relation in the CGM traced by MgII absorbers deviates from the Milky Way relationship by $5.3\sigma$.

In Figure~\ref{EvsO}, we also overplot the 3$\sigma$ upper limits of DIB$\lambda4430$ absorption strengths in damped Lyman alpha systems (DLAs) from \citet{Lawton2006} (purple data points). The authors searched for DIBs in seven DLAs and detected DIBs in only one system having the highest reddening and metallicity. They also reached to a similar conclusion that the DIB-E(B-V) relation in the DLAs is deviated from the Milky Way relation. This trend is also observed in the Large and Small Magellanic Clouds, showing weaker DIB absorption lines than values based on Milky Way DIB - dust/gas relation \citep{Welty2006}.

Our results indicate that with the same E(B-V) value, the abundances of the DIB$\lambda4430$ carrier in the ISM and in the CGM are significantly different, indicating that factors associated with different environments play a key role in regulating the abundance of the DIB$\lambda4430$ carrier. Here we discuss possible factors for the different strengths of DIB$\lambda4430$ in the MW and in MgII absorbers:
\begin{itemize}
    \item \textbf{Production of the DIB carrier:} One possibility is that a certain environment is required to form the DIB$\lambda4430$ carrier; the Milky Way has such an environment while the circumgalacitc gas traced by MgII absorbers and galaxies producing the gas do not have such an environment. This might possibly link to the $\rm 2175 \AA$ bump in the extinction curve. Both the Milky Way \citep[e.g.,][]{Fitzpatrick2007} and the DLA with DIB$\lambda4430$ detected \citep{Junkkarinen2004} have the $\rm 2175 \AA$ bump in their extinction curves, while SMC and MgII absorbers \citep{York2006, Menard2012} have similar extinction curves without the bump and weaker DIB absorption strength. We note that while the correlation between $\rm 2175 \AA$ bump and DIB$\lambda4430$ is still under debate \citep[e.g.,][]{Xiang2017, Cox2007, Iglesias2007}, if the correlation is further confirmed, it indicates that the environments under certain conditions can grow materials being responsible for $\rm 2175 \AA$ extinction bump and DIB$\lambda4430$.

    \item \textbf{Destruction of the DIB carrier:} Given that MgII absorbers trace the circumgalactic gas clouds with part of the materials expected to be ejected from galaxies through feedback mechanisms \citep[e.g.,][]{Bordoloi2011, Lan2018}, it is possible that with the energy from stellar explosions, the carrier of DIB$\lambda4430$ is ionized and/or destroyed during the transportation process, similar to the dust destruction due to supernovae \citep[e.g.,][]{Priestley2021}. The results of \citet{Milisavljevic2014} are consistent with this scenario. The authors observed the spectra of Type Ic supernova SN 2012ap as a function of time and found that the absorption strength of DIB$\lambda4430$ decreases by a factor of two over a 10-day period, an evidence of the DIB$\lambda4430$ carrier being ionized by the radiation produced by the supernova. This indicates that the DIB$\lambda4430$ carrier tends to be less abundant in environments with strong radiation fields. The authors also found that DIB$\lambda5780$ shows a different trend. Its absorption strength increases by $\sim 20\%$ over a similar timescale of the change of DIB$\lambda4430$. The authors argued that the carrier of DIB$\lambda5780$ might be photo-products of the carrier of DIB$\lambda4430$. In addition, with a relatively lower gas density in the CGM \citep[e.g.,][]{Tumlinson2017}, materials in the CGM are expected to be in higher ionization states than gas and molecules in the ISM. Therefore, if the carrier of DIB$\lambda4430$ is neutral or in lower ionization state, it is possible that they are in a higher ionization state in the CGM.  
    This can be another factor for the non-detection of DIB$\lambda4430$ in MgII absorbers. 
\end{itemize}

\section{Conclusions}
\label{section_4}
In this work, we performed a deep search for DIB$\lambda4430$, one of the strongest DIBs observed in the Milky Way, in the extragalactic circumgalactic gas traced by MgII absorption lines at $0.4<z<1.0$. To this end, we obtained high S/N composite spectra using $\sim 60,000$ SDSS absorption line spectra from which we measured the DIB$\lambda4430$ rest equivalent widths. Our results are summarized as follows:
\begin{itemize}
    \item The composite spectra enable us to 
     probe the DIB$\lambda4430$ rest equivalent widths as a function of MgII rest equivalent widths with absorption strength uncertainty being a few $\rm m\AA$. However, no DIB$\lambda4430$ absorption signals are detected.

     \item Adopting the MgII rest equivalent width and dust E(B-V) relation from the literature, we compare the relation between DIB$\lambda4430$ absorption strength and E(B-V) in the CGM and the relation obtained in the ISM of the Milky Way. We find that with the same E(B-V), the DIB$\lambda4430$ absorption is lower than the DIB$\lambda4430$ absorption obtained in the Milky Way, yielding $\sim 5\sigma$ deviation between the DIB$\lambda4430$-E(B-V) relation in the CGM and the relation in the ISM. 

    \item We also produce synthetic spectra by adding DIB absorption lines with strengths, following the MW DIB$\lambda4430$-E(B-V) relation, in the quasar spectra, perform the analysis procedure, and demonstrate that the input signals can be recovered and detected with the dataset. 
\end{itemize}
Our results show that with the same reddening values, the abundance of the DIB$\lambda4430$ carrier in the CGM environment is lower than the abundance of the DIB$\lambda4430$ carrier in the ISM environment, indicating that the environmental factors, such as gas density, radiation field, and transportation of materials from the ISM to the CGM, play a key role in regulating the amount of DIB carriers. 

This work also demonstrates that probing the DIBs in different environments provides useful information on understanding the nature of DIB carriers. Ongoing and upcoming spectroscopic surveys will offer rich datasets enabling the explorations of DIBs in different environments with unprecedented precision. For example, the Dark Energy Spectroscopic Instrument (DESI) \citep{Levi2013,desi2022} will collect a few million quasar spectra and detect hundred thousands of metal absorbers \citep[e.g.,][]{Napolitano2023}, which will increase the number of absorption lines by a factor of 4-5 for a similar study of probing DIBs in the CGM environment. In addition, spatial-resolved spectroscopic observations with integral field units for galaxies can potentially be used to map out the DIB distribution in nearby extragalactic galaxies and explore the properties of galaxies and DIBs \citep[e.g.,][]{Musedib}. 
Finally, the number of spectra of stars and extragalactic sources from sky surveys, e.g., DESI, SDSS-V \citep{Kollmeier2017}, 4MOST \citep{4most19}, PFS \citep{Takada2014}, continues to grow. The distributions of DIBs in various ISM environments in the Milky Way can be mapped in greater detail, revealing the underlying mechanisms that regulate the production and destruction of DIB carriers.

\vspace{0.5cm}

C.Y.C and T.W.L are supported by the National Science and Technology Council (MOST 111-2112-M-002-015-MY3, NSTC 113-2112-M-002-028-MY3), the Yushan Fellow Program by the Ministry of Education (MOE) (NTU-110VV007, NTU-111V1007-2, NTU-112V1007-3, NTU-113V1007-4), National Taiwan University research grant (NTU-CC-111L894806, NTU-CC-112L893606, NTU-CC-113L891806, NTU-111L7318, NTU-112L7302). C.Y.C acknowledges the support of the National Science and Technology Council (111-2813-C-002-010-M).

Funding for the Sloan Digital Sky Survey IV has been provided by the Alfred P. Sloan Foundation, the U.S. Department of Energy Office of Science, and the Participating Institutions. SDSS acknowledges support and resources from the Center for High-Performance Computing at the University of Utah. The SDSS web site is www.sdss4.org.

SDSS is managed by the Astrophysical Research Consortium for the Participating Institutions of the SDSS Collaboration including the Brazilian Participation Group, the Carnegie Institution for Science, Carnegie Mellon University, Center for Astrophysics | Harvard \& Smithsonian (CfA), the Chilean Participation Group, the French Participation Group, Instituto de Astrofísica de Canarias, The Johns Hopkins University, Kavli Institute for the Physics and Mathematics of the Universe (IPMU) / University of Tokyo, the Korean Participation Group, Lawrence Berkeley National Laboratory, Leibniz Institut für Astrophysik Potsdam (AIP), Max-Planck-Institut für Astronomie (MPIA Heidelberg), Max-Planck-Institut für Astrophysik (MPA Garching), Max-Planck-Institut für Extraterrestrische Physik (MPE), National Astronomical Observatories of China, New Mexico State University, New York University, University of Notre Dame, Observatório Nacional / MCTI, The Ohio State University, Pennsylvania State University, Shanghai Astronomical Observatory, United Kingdom Participation Group, Universidad Nacional Autónoma de México, University of Arizona, University of Colorado Boulder, University of Oxford, University of Portsmouth, University of Utah, University of Virginia, University of Washington, University of Wisconsin, Vanderbilt University, and Yale University.

\end{document}